\pageno=1                                      
%
%
%
\font\ninerm=cmr9
\font\eightrm=cmr8
\font\sixrm=cmr6
\font\ninei=cmmi9
\font\eighti=cmmi8
\font\sixi=cmmi6
\skewchar\ninei='177 \skewchar\eighti='177 \skewchar\sixi='177
\font\ninesy=cmsy9
\font\eightsy=cmsy8
\font\sixsy=cmsy6
\skewchar\ninesy='60 \skewchar\eightsy='60 \skewchar\sixsy='60

\font\ninebf=cmbx9
\font\eightbf=cmbx8
\font\sixbf=cmbx6
\font\ninett=cmtt9
\font\eighttt=cmtt8
\hyphenchar\tentt=-1 
\hyphenchar\ninett=-1
\hyphenchar\eighttt=-1
\font\ninesl=cmsl9
\font\eightsl=cmsl8
\font\nineit=cmti9
\font\eightit=cmti8
\newskip\ttglue
\def\tenpoint{\def\rm{\fam0\tenrm}%
  \textfont0=\tenrm \scriptfont0=\sevenrm \scriptscriptfont0=\fiverm
  \textfont1=\teni \scriptfont1=\seveni \scriptscriptfont1=\fivei
  \textfont2=\tensy \scriptfont2=\sevensy \scriptscriptfont2=\fivesy
  \textfont3=\tenex \scriptfont3=\tenex \scriptscriptfont3=\tenex
  \def\it{\fam\itfam\tenit}%
  \textfont\itfam=\tenit
  \def\sl{\fam\slfam\tensl}%
  \textfont\slfam=\tensl
  \def\bf{\fam\bffam\tenbf}%
  \textfont\bffam=\tenbf \scriptfont\bffam=\sevenbf
   \scriptscriptfont\bffam=\fivebf
  \def\tt{\fam\ttfam\tentt}%
  \textfont\ttfam=\tentt
  \tt \ttglue=.5em plus.25em minus.15em
  \normalbaselineskip=12pt
  \let\sc=\eightrm
  \let\big=\tenbig
  \setbox\strutbox=\hbox{\vrule height8.5pt depth3.5pt width0pt}%
  \normalbaselines\rm}
\def\ninepoint{\def\rm{\fam0\ninerm}%
  \textfont0=\ninerm \scriptfont0=\sixrm \scriptscriptfont0=\fiverm
  \textfont1=\ninei \scriptfont1=\sixi \scriptscriptfont1=\fivei
  \textfont2=\ninesy \scriptfont2=\sixsy \scriptscriptfont2=\fivesy
  \textfont3=\tenex \scriptfont3=\tenex \scriptscriptfont3=\tenex
  \def\it{\fam\itfam\nineit}%
  \textfont\itfam=\nineit
  \def\sl{\fam\slfam\ninesl}%
  \textfont\slfam=\ninesl
  \def\bf{\fam\bffam\ninebf}%
  \textfont\bffam=\ninebf \scriptfont\bffam=\sixbf
   \scriptscriptfont\bffam=\fivebf
  \def\tt{\fam\ttfam\ninett}%
  \textfont\ttfam=\ninett
  \tt \ttglue=.5em plus.25em minus.15em
  \normalbaselineskip=10pt 
  \let\sc=\sevenrm
  \let\big=\ninebig
  \setbox\strutbox=\hbox{\vrule height8pt depth3pt width0pt}%
  \normalbaselines\rm}
\def\eightpoint{\def\rm{\fam0\eightrm}%
  \textfont0=\eightrm \scriptfont0=\sixrm \scriptscriptfont0=\fiverm
  \textfont1=\eighti \scriptfont1=\sixi \scriptscriptfont1=\fivei
  \textfont2=\eightsy \scriptfont2=\sixsy \scriptscriptfont2=\fivesy
  \textfont3=\tenex \scriptfont3=\tenex \scriptscriptfont3=\tenex
  \def\it{\fam\itfam\eightit}%
  \textfont\itfam=\eightit
  \def\sl{\fam\slfam\eightsl}%
  \textfont\slfam=\eightsl
  \def\bf{\fam\bffam\eightbf}%
  \textfont\bffam=\eightbf \scriptfont\bffam=\sixbf
   \scriptscriptfont\bffam=\fivebf
  \def\tt{\fam\ttfam\eighttt}%
  \textfont\ttfam=\eighttt
  \tt \ttglue=.5em plus.25em minus.15em
  \normalbaselineskip=9pt
  \let\sc=\sixrm
  \let\big=\eightbig
  \setbox\strutbox=\hbox{\vrule height7pt depth2pt width0pt}%
  \normalbaselines\rm}
%
\def\headtype{\ninepoint}                 
\def\abstracttype{\ninepoint}             
\def\captiontype{\ninepoint}              
\def\footnotetype{\ninepoint}             
\def\refit{\it}                           
\font\chaptitle=cmr10 at 11pt             
\rm                                       

%
%
\parindent=0.25in                         
\parskip=0pt                              
\baselineskip=12pt                        
\hsize=4.25truein                         
\vsize=7.445truein                        
\hoffset=1in                              
\voffset=-0.5in                           

\newskip\sectionskipamount                
\newskip\aftermainskipamount              
\newskip\subsecskipamount                 
\newskip\firstpageskipamount              
\newskip\capskipamount                    
\newskip\ackskipamount                    
\sectionskipamount=0.2in plus 0.09in
\aftermainskipamount=6pt plus 6pt         
\subsecskipamount=0.1in plus 0.04in
\firstpageskipamount=3pc
\capskipamount=0.1in
\ackskipamount=0.15in
\def\sectionskip{\vskip\sectionskipamount}
\def\aftermainskip{\vskip\aftermainskipamount}
\def\subsecskip{\vskip\subsecskipamount} 
\def\firstpageskip{\vskip\firstpageskipamount}
\def\capskip{\hskip\capskipamount}

%
%
\nopagenumbers                            
\newcount\firstpageno                     
\firstpageno=\pageno                      
\newcount\chapno                          

\def\rightheadline{\headtype\phantom{\folio}\hfil\runningtitletext\hfil\folio}
\def\leftheadline{\headtype\folio\hfil\runningauthortext\hfil\phantom{\folio}}
\headline={\ifnum\pageno=\firstpageno\hfil
           \else
              \ifdim\ht\topins=\vsize           
                 \ifdim\dp\topins=1sp \hfil     
                 \else
                     \ifodd\pageno\rightheadline\else\leftheadline\fi
                 \fi
              \else
                 \ifodd\pageno\rightheadline\else\leftheadline\fi
              \fi
           \fi}

\def\bottomnumber{\hss\tenrm[\folio]\hss}
\footline={\ifnum\pageno=\firstpageno\bottomnumber\else\hfil\fi}

%
%
%
%
\outer\def\mainsection#1
    {\vskip 0pt plus\smallskipamount\sectionskip
     \message{#1}\vbox{\noindent{\bf#1}}\nobreak\aftermainskip\noindent}
 
\outer\def\subsection#1
    {\vskip 0pt plus\smallskipamount\subsecskip
     \message{#1}\vbox{\noindent{\bf#1}}\nobreak\smallskip\nobreak\noindent}
 
\def\backup{\nobreak\vskip-\baselineskip\nobreak\vskip-\subsecskipamount\nobreak
}

\def\title#1{{\chaptitle\leftline{#1}}}
\def\name#1{\leftline{#1}}
\def\affiliation#1{\leftline{\it #1}}
\def\abstract#1{{\abstracttype \noindent #1 \smallskip\vskip .1in}}
\def\ref{\noindent \parshape2 0truein 4.25truein 0.25truein 4truein}
\def\caption{\noindent \captiontype
             \parshape=2 0truein 4.25truein .125truein 4.125truein}

\def\footnote#1{\edef\fspafac{\spacefactor\the\spacefactor}#1\fspafac
      \insert\footins\bgroup\footnotetype
      \interlinepenalty100 \let\par=\endgraf
        \leftskip=0pt \rightskip=0pt
        \splittopskip=10pt plus 1pt minus 1pt \floatingpenalty=20000
        \textindent{#1}\bgroup\strut\aftergroup\strut\egroup\let\next}
\skip\footins=12pt plus 2pt minus 4pt 
\dimen\footins=30pc 

%
%

\def\@{\spacefactor 1000}

\def\,{\pcomma} 
\def\pcomma{\relax\ifmmode\mskip\thinmuskip\else\thinspace\fi}

\def\oversim#1#2{\lower0.5ex\vbox{\baselineskip=0pt\lineskip=0.2ex
     \ialign{$\mathsurround=0pt #1\hfil##\hfil$\crcr#2\crcr\sim\crcr}}}


\def\wig#1{\mathrel{\hbox{\hbox to 0pt{%
          \lower.5ex\hbox{$\sim$}\hss}\raise.4ex\hbox{$#1$}}}}

\def\v1n{{\cal U}^N_1}

\def\sss{\scriptscriptstyle}
\def\phih2h2{\phi_{\sss {\rm H_2-H_2}}}
\def\phh2{\phi_{\sss {\rm H-H_2}}}

\def\etal{{ et~al.}\ }
\def\Teff{{\rm T}_{\rm eff}}
\def\sqr#1#2{{\vcenter{\vbox{\hrule height.#2pt
  \hbox{\vrule width.#2pt height#1pt \kern#1pt
  \vrule width.#2pt}
  \hrule height.#2pt}}}}

\def\wig#1{\mathrel{\hbox{\hbox to 0pt{%
          \lower.5ex\hbox{$\sim$}\hss}\raise.4ex\hbox{$#1$}}}}

\def\mo{$M_\odot$}

\def\lo{$L_\odot\,$}
\def\mj{$M_{\rm J}\,$}

\def\mstar{$M_{\ast}$}
\def\rj{R$_{\rm J}\ $}
\def\undertext#1{$\underline{\smash{\hbox{#1}}}$}
\def\sles{\lower2pt\hbox{$\buildrel {\scriptstyle <}
   \over {\scriptstyle\sim}$}}
\def\sgreat{\lower2pt\hbox{$\buildrel {\scriptstyle >}
   \over {\scriptstyle\sim}$}}

\def\runningtitletext{}
\def\runningauthortext{}

\null
\firstpageskip

{\baselineskip=14pt
\title{NEW IDEAS IN THE THEORY OF} 
\title{EXTRASOLAR GIANT PLANETS AND BROWN DWARFS}
}

\vskip .3truein
\name{ADAM BURROWS, BILL HUBBARD, and JONATHAN LUNINE}
\affiliation{University of Arizona}
\vskip .2truein
\name{MARK MARLEY}
\affiliation{New Mexico State University}
\vskip .1truein
\leftline{and}
\vskip .1truein
\name{DIDIER SAUMON}
\affiliation{Vanderbilt University}
\vskip .3truein


\mainsection{{I}.~~INTRODUCTION}
\backup
The study of extrasolar giant planets (EGPs\footnote*{We use this
shorthand for \undertext{E}xtrasolar \undertext{G}iant \undertext{P}lanet, 
but the terms ``exoplanet'' or ``super--jupiter''
are equally appropriate.}) and brown dwarfs via reflex stellar motion, broad--band photometry,
and spectroscopy has finally come into its own.
Doppler spectroscopy alone has revealed about 15 objects
in the giant planet/brown dwarf regime, including companions to $\tau$ Boo, 51 Peg, $\upsilon$ And,
55 Cnc, $\rho$ CrB, 70 Vir, 16 Cyg B, and 47 UMa (Noyes \etal\ 1997; Butler \etal\ 1997; Cochran \etal\ 1997;
Marcy \& Butler 1996; Butler \& Marcy 1996; Mayor \& Queloz 1995;
Latham \etal\ 1989).  

The direct detection of Gl229 B
(Oppenheimer \etal\  1995; Nakajima \etal\ 1995; Matthews \etal\ 1996; Geballe \etal\ 1996;
Marley \etal\ 1996; Allard \etal\ 1996; Tsuji \etal\ 1996) was a milestone since Gl229 B displays methane spectral features
and low surface fluxes that are unique to objects with effective temperatures (in this case, T$_{\rm eff}$$\sim$950 K)
below those at the solar--metallicity main sequence edge ($\sim$1750 K, Burrows \etal 1993).
In addition, the almost complete absence of spectral signatures of metal oxides and hydrides (such as
TiO, VO, FeH, and CaH) is in keeping with theoretical predictions that these species are depleted
in the atmospheres of all but the youngest (hence, hottest) substellar objects
and are sequestered in condensed form below the photosphere (Lunine \etal\ 1989; Marley \etal\ 1996).
The wide range in mass
and period, as well as the proximity of many of the planets/brown dwarfs
to their primaries, was not anticipated by most planetary scientists.
Though the technique of Doppler spectroscopy used to find these companions
selects for massive, nearby
objects, their variety and existence is a challenge to conventional
theory.
Since direct detection is now feasible, and has been
demonstrated by the recent acquisition of Gl229 B, it is crucial for the future of extrasolar giant planet
searches that the spectra, colors, evolution, and physical structure of objects from Saturn's mass (0.3 \mj) to 70 \mj
be theoretically investigated.

\mainsection{{I}{I}.~~EARLY CALCULATIONS OF THE EVOLUTION AND STRUCTURE OF EXTRASOLAR GIANT PLANETS}

\subsection{{A}. Gray Models}

EGPs radiate in the optical by reflection and in the
infrared by the thermal emission of both absorbed stellar light and the planet's
own internal energy.
In Burrows \etal\ (1995) and Saumon \etal\ (1996), the EGPs were assumed to be fully convective at all times.
We included the effects of ``insolation'' by a central star of mass \mstar$\,$ and considered
semi-major axes ($a$) between 2.5 A.U. and 20 A.U.  Giant
planets may form preferentially near 5 A.U. (Boss 1995), but as the new data dramatically affirm,
a broad range of semi--major axes can not be excluded.
In these calculations, we assumed that the Bond albedo of an 
EGP is that of Jupiter ($\sim$0.35).
For the Burrows \etal\ (1995) study, we evolved EGPs with masses from 0.3 \mj $\,$(the mass of Saturn)
through 15 \mj .
Whether a 15 \mj object is a planet or a brown dwarf is largely a semantic issue, though one might
distinguish gas giants and brown dwarfs by their
mode of formation (e.g., in a disk or ``directly'').   

If 51 Peg b is a gas giant, its radius is only 1.2$\,$R$_{\rm J}$ and its
luminosity is about $3.5\times 10^{-5}\,L_\odot$.  This bolometric luminosity is more than $1.5 \times 10^4$
times the present luminosity of Jupiter and only a factor of two below that at the edge of the main sequence.
The radiative region encompasses
the outer 0.03\% in mass, and 3.5\% in radius.
The study by Guillot \etal\ (1996) demonstrated that 51 Peg b is well within its Roche lobe
and is not experiencing significant photoevaporation.  Its deep potential well ensures that,
even so close to its parent, 51 Peg b is {\bf stable}.
If 51 Peg b were formed beyond 1 A.U. and moved inward on a timescale greater than $\sim 10^{8}$ years, it would
closely follow a $R_{\rm p} \sim$ \rj trajectory to its equilibrium position.

\subsection{{B}. Non--Gray Models}

However, to credibly estimate the infrared band fluxes and improve 
upon the black body assumption made in Burrows \etal\ (1995)
and  Saumon \etal\ (1996), we have
recently performed {\bf non-gray} simulations at solar--metallicity of the evolution, spectra,
and colors of isolated EGP/brown dwarfs down to T$_{\rm eff}$s of 100 K (Burrows \etal\ 1997).
%
%
Figure 1 portrays the luminosity versus time for objects from Saturn's mass (0.3 \mj) to 0.2 \mo$\,$ for this model suite.
The early plateaux between 10$^6$ years and 10$^8$ years are due to deuterium burning, where
the initial deuterium mass fraction was taken to be 2$\times$10$^{-5}$.  Deuterium burning occurs earlier,
is quicker, and is at higher luminosity for the more massive models, but can take as long
as 10$^{8}$ years for a 15 \mj object.  The mass below which less than 50\% of the ``primordial''
deuterium is burnt is $\sim$13 \mj (Burrows \etal\ 1995).  On this figure, we have arbitrarily
classed as ``planets'' those objects that do not burn deuterium and as ``brown dwarfs'' those that do burn deuterium,
but not light hydrogen.  While this distinction is physically motivated, we do not
advocate abandoning the definition based on origin.  Nevertheless, the separation
into M dwarfs, ``brown dwarfs'', and giant ``planets'' is useful for parsing by eye the information in the figure.

In Fig. 1, the bumps between 10$^{-4}$ \lo and 10$^{-3}$ \lo and between 10$^{8}$ and 10$^{9}$ years,
seen on the cooling curves of objects from 0.03 \mo to 0.08 \mo, 
are due to silicate and iron grain formation.   These effects,
first pointed out by Lunine \etal\ (1989), occur for T$_{\rm eff}$s between 2500 K and 1300 K.
The presence of grains affects the precise mass and luminosity at the edge of the main sequence.
Since grain and cloud models are problematic, there still remains much to learn concerning
their role and how to model them (Lunine \etal\ 1989; Allard \etal\ 1997; Tsuji \etal 1996).

\mainsection{{I}{I}{I}. NEW INSIGHTS}

The studies of Burrows \etal\ (1997) and Marley \etal\ (1996) reveal major new aspects of
EGP/brown dwarf atmospheres that bear listing and that uniquely
characterize them.
Below T$_{\rm eff}$s of 1300 K, the dominant equilibrium carbon molecule is CH$_4$, not CO,
and below 600 K the dominant nitrogen molecule is NH$_3$, not N$_2$ (Fegley \& Lodders 1996).
In objects with T$_{\rm eff}$ $\le$ 1300 K, the major opacity sources are H$_2$, H$_2$O, CH$_4$, and NH$_3$.
For T$_{\rm eff}$s below $\sim$400 K, water clouds form at or above the photosphere (defined where T = T$_{\rm eff}$)
and for T$_{\rm eff}$s below 200 K, ammonia clouds form (viz., Jupiter).  Collision--induced absorption
of H$_2$ partially suppresses emissions longward of $\sim$10 $\,\mu$m.  The holes in the opacity
spectrum of H$_2$O that define the classic telluric IR bands also regulate much of the emission from
EGP/brown dwarfs in the near infrared.  Importantly, the windows in H$_2$O and the suppression by H$_2$ conspire to
force flux to the blue for a given T$_{\rm eff}$.
The upshot is an exotic spectrum enhanced relative to the black body value
in the $J$ and $H$ bands ($\sim$1.2 $\,\mu$m\ and $\sim$1.6 $\,\mu$m, respectively) 
by as much as {\it two} to {\it ten} orders of magnitude,
depending upon T$_{\rm eff}$.
The enhancement at 5 $\,\mu$m\ for a 1 Gyr old, 1 \mj extrasolar planet is by four orders
of magnitude.
As T$_{\rm eff}$ decreases below $\sim$1000 K, the flux in the $M$ band ($\sim$5 $\,\mu$m)
is progressively enhanced relative to the black body value.
While at 1000 K there is no enhancement, at 200 K it is near 10$^5$.   The $J$, $H$, and $M$ bands
are the premier bands in which to search for cold substellar objects.   The $Z$ band ($\sim$1.05 $\,\mu$m) is also
in excess of the black body value over this T$_{\rm eff}$ range. 
Even though $K$ band ($\sim$2.2 $\,\mu$m) fluxes are generally higher
than black body values,  H$_2$ and CH$_4$ absorption features in the $K$ band decrease its importance
{\it relative} to $J$ and $H$.  As a consequence of the increase of atmospheric
pressure with decreasing T$_{\rm eff}$, the anomalously blue $J-K$ and $H-K$
colors get {\it bluer},
not redder.
The $K$ and $J$ versus $J-K$ infrared H--R diagrams loop back to the blue
below the edge
of the main sequence and are not continuations of the M dwarf sequence into the red.
The difference between the black body curves and the model curves is between 3 and 10 magnitudes
for $J$ versus $J-K$, more for $K$ versus $J-K$.  Gl229 B fits nicely among these theoretical isochrones.
The suppression of $K$ by H$_2$ and CH$_4$ features is largely responsible for this anomalous blueward
trend with decreasing mass and T$_{\rm eff}$.

\mainsection{{I}{V}.~~ATMOSPHERIC COMPOSITIONS OF EGPS, CONDENSATION, AND CLOUDS}
\backup

The molecular compositions of the exotic, low--ionization atmospheres
of EGPs and brown dwarfs can serve as diagnostics of temperature, mass, and elemental abundance and can help define
a spectral sequence, just as the presence or absence of spectral features associated with various
ionization states of dominant, or spectroscopically active,
atoms and simple molecules does for M through O stars.
However, the multiplicity of molecules that appear in their atmospheres lends an additional complexity to
the study of substellar mass objects that is both helpfully diagnostic and confusing.
Nowhere is the latter more apparent than in the appearance
at low temperatures of refractory grains and clouds.
These condensed species can contribute significant opacity
and can alter an atmosphere's temperature/pressure
profile and its albedo.  Grain and cloud droplet opacities
depend upon the particle size and shape distribution
and these are intertwined with the meteorology (convection) in complex ways.
Furthermore, condensed species can rain out and deplete the upper atmosphere of heavy elements, thereby
changing the composition and the observed spectrum.
In brown dwarf and EGP atmospheres, abundance and temperature/pressure profiles,
particle properties, spectra, and meteorology are inextricably linked.

The formation of refractory silicate grains
below 2500 K was already shown by Lunine \etal (1989) and Burrows \etal (1989)
to influence the evolution of late M dwarfs and young brown dwarfs through their ``Mie'' opacity.
The blanketing effect they provide lowers the effective temperature (T$_{\rm eff}$)
and luminosity (L) of the main sequence edge mass from about 2000 K and $10^{-4}$ \lo
to about 1750 K and $6\times10^{-5}$ \lo, an effect recently verified by Chabrier \etal (1998).
In addition, grain opacity slightly delays the cooling of older brown dwarfs, imprinting a slight
bump on their luminosity/age trajectories (see Fig. 1).
The presence of grains in late M dwarf spectra was invoked
to explain the weakening of the TiO bands and the shallowing of their H$_2$O troughs in the
near infrared (Tsuji \etal 1996; Jones \& Tsuji 1997).  Tsuji
and collaborators concluded that titanium was being depleted into refractories,
a conclusion with which we agree.

With a T$_{\rm eff}$ of $\sim$950 K, a luminosity below  $10^{-5}$ \lo, and spectra or photometry
from the $R$ band through 5 $\,\mu$m, Gl229 B hints at or exemplifies all of the unique characteristics
of the family: metal (Fe, Ti, V, Ca, Mg, Al, Si) depletions,
the dominance of H$_2$O vapor, the appearance of CH$_4$ and alkali metals, and the signatures of clouds.
Clouds of low--temperature condensible species above the photosphere are the most
natural explanation for the steep drop below 1 $\,\mu$m in the Keck spectra between 0.83 $\,\mu$m
and 1 $\,\mu$m (Oppenheimer \etal 1998).  These clouds might not be made up of the classic
silicate refractories formed at much higher temperatures, since these species may
have rained out.  From simple Mie theory, their mean particle size must be
small ($\sim$ 0.1 $\,\mu$m) in order to influence the ``optical'' without much perturbing the near infrared.
In addition, such a population of small droplets can help explain why
Gl229 B's near--infrared troughs at 1.8 $\,\mu$m and 3.0 $\,\mu$m are not as deep as theory would otherwise
have predicted.  Just as Tsuji and collaborators have shown that silicate grains at higher temperatures
can shallow out the H$_2$O troughs, so to can species that condense at lower temperatures
($\le$ 1000 K ?) explain the shallower--than--predicted Gl229 B H$_2$O troughs.  What those species might be
can be illuminated by chemical abundance studies (Burrows \& Sharp 1998).
Note that a cloud grammage in these small--radius, low--temperature 
refractories of only $\sim 10^{-5}$ gm cm$^{-2}$
would be adequate to explain the anomalies.

\subsection{A.~~Abundance and Condensate Results}

At temperatures of $\sim$1500--2000 K, 
the standard refractories, such as the silicates, spinel, and iron,
condense out into grain clouds which by their large opacity
lower the T$_{\rm eff}$ and luminosity of the
main sequence edge (Burrows \etal 1993; Burrows \& Sharp 1998) and alter
in detectable ways the spectra of objects
around the transition mass (Jones \& Tsuji 1997).  As T$_{\rm eff}$ decreases
below that at the stellar edge, the classical refractories
are buried progressively deeper below the photosphere and less refractory
condensates and gas--phase molecules come to dominate
(Marley \etal 1996; Burrows \etal 1997).

Below temperatures of
$\sim$1500 K, the calculations of Burrows \& Sharp (1998) demonstrate and confirm that
the alkali metals, which are not as refractory
as Fe, Al, Ca, Ti, V, and Mg, emerge as important
atmospheric and spectral constituents.  At still lower temperatures,
chlorides and sulfides appear, some of which will condense in the cooler
upper atmosphere and form clouds that will affect emergent spectra and
albedos.  Cloud decks of many different compositions at many different
temperature levels are expected, depending upon T$_{\rm eff}$ (and weakly
upon gravity).  Clouds of chlorides and sulfides
at temperature levels below $\sim$1000 K may be responsible
for the steeper slope observed in the spectrum of Gl229 B
at the shorter wavelengths (Oppenheimer \etal 1998).
At slightly higher temperatures, MnS, ZnS, NaAlSi$_3$O$_8$,
KAlSi$_3$O$_8$, V$_2$O$_3$,
and MgTi$_2$O$_5$ may play a role, but only if
their constituents are not scavenged into more
refractory compounds and rained out deeper down.

As T$_{\rm eff}$ decreases (either as a given mass cools or, for a given age,
as we study objects with lower masses), the major atmospheric constituents of
brown dwarfs and EGPs change. This change is reflected in
which spectral features are most prominent
and in the albedos of substellar objects near their primaries.
Equilibrium chemical sequences
are predominantly a function of temperature and can help to define a spectral sequence
for substellar objects from the main sequence edge near 2000 K to EGPs with T$_{\rm eff}$s of a few
hundred Kelvin.  The appearance and disappearance of various molecules and refractories delineates
an effective temperature sequence and the new proposed ``L'' dwarf
spectral classification (Kirkpatrick \etal 1998) may correspond to a subset of such a compositional
sequence.
Very crudely, the ``L'' spectral type suggested by Kirkpatrick \etal (1998) would
correspond to T$_{\rm eff}$s between about $\sim$1500 K and $\sim$2200 K.
All but the youngest and most massive brown dwarfs and
only the very youngest EGPs could have this proposed spectral designation.  Most brown dwarfs and EGPs
will be of an even later spectral type, yet to be coined, a spectral type that would include Gl229 B.

\subsection{{B}.~~Clouds}

Cloud formation refers to the production of solid or liquid particles in the atmosphere of a brown
dwarf or extrasolar planet. Cloud formation in the Earth's troposphere is a complex phenomenon
involving a single species (water), two phases, multiple particle size distributions, incompletely--known
particle properties and myriad cloud morphologies. The range of radiative properties of
terrestrial clouds is commensurately broad, and different types of clouds can serve to either warm
or cool the surface of the Earth. Much of the complexity lies in the heterogeneous distribution of
such clouds in vertical and horizontal directions; the problem of treatment of the radiative transfer
of broken clouds remains a difficult one (Goody \& Yung 1989). Furthermore, cloud formation
itself contributes latent heat to an atmosphere and, hence, can destabilize a radiative region,
requiring treatment of the heat transfer by moist convection (Emanuel 1994). Cloud--forming
species are known or suspected to be present in eight of the nine planets of the solar system, as
well as two of the moons of the outer solar system (Titan and Triton).

The giant planets of our solar system remain the best guide to the study of clouds in brown dwarfs
and extrasolar giant planets, because the background gas is hydrogen--helium in all cases, and the
atmospheres are not thin layers atop a solid surface. Principal cloud forming species in the upper
Jovian and Saturnian atmosphere are ammonia, water, and possibly ammonium hydrosulfide;
in Uranus and Neptune they are methane and either ammonium hydrosulfide or hydrogen
sulfide (Baines \etal\ 1995).  However, a number of minor species may participate in cloud
formation, and stratospheric photochemical hazes are generated by the action of solar
ultraviolet radiation. In situ observations of Jupiter's cloud structure by the Galileo entry probe
revealed the extent of heterogeneity across the surface of a gas giant planet; the probe failed to
detect anything but very a tenuous and narrow cloud layer (Ragent \etal\ 1995). Since ample
visual and spectroscopic evidence exist for widespread clouds on Jupiter (Carlson \etal
1995), the Galileo probe must have found a region of unusual dryness. This notion is
supported by the observation that the probe fell into a ``hot spot,'' a region of unusually high 
5--micron emission (Orton \etal\ 1996).

Modeling a potentially broad range of compositions and properties from the small number of
detailed observations of clouds available to date is daunting. One strategy is to consider 
end-member cloud models, characterized by simple condensation in a radiative atmosphere on the one
hand, and by large scale vertical motions driven by thermal convection or other processes on the
other (Lunine \etal 1986; 1989). Such models can be reduced to a few parameters which can
then be constrained, albeit weakly, with spectral observations of the atmosphere. In the radiative
case the cloud mass density versus altitude is determined by the saturation vapor pressure and an
assumed value for the meteorological supersaturation, which determines the condensable available
to form clouds, but is determined for the Earth empirically (Emanuel 1994). This factor is difficult
to determine a priori for an object in which the clouds cannot be directly studied, but can be
crudely estimated (Rossow 1978). A simple prescription for the cloud number density for a
convectively unstable brown dwarf atmosphere is given by Lunine \etal\ (1989). In each case the
mean particle size can be determined semi--analytically (Rossow 1978) or by use of a numerical
particle growth model (Yair \etal\ 1995). The problem with the former approach is that the mean
particle size can be significantly underestimated, based on terrestrial experience, and a size
distribution is not obtained. The numerical approach suffers from the problem of poor specificity
of input parameters when applied to extrasolar planets and brown dwarfs.

Spectroscopy of Gliese 229 B suggests that grain formation may be occurring in its atmosphere.
Grains both alter the spectral contrast of molecular lines and, through
condensation, remove gas phase species that may be directly responsible for various absorption
features. Because of condensation, silicate or iron features in the spectrum may disappear at
around or above 1000 K, with gaseous water bands doing the same around 400 K. The effects of clouds on
the albedo or reflectivity of a brown dwarf or extrasolar planet are also large. 
Modeling by Marley \etal\ (1998) suggests large variations (factors of two or more) in the albedo of a
brown dwarf or extrasolar planet, depending upon the type of cloud (large--particle convective
versus small-particle laminar clouds). The four giant planets all have surface reflectivities
influenced by a combination of cloud and gas opacity sources, to varying extents. 

Further progress in quantifying the extent of the effects of clouds on brown dwarf atmospheres
will require more realistic models of cloud formation, transport and growth of particles, data on
indices of refraction of cloud particles, and incorporation of moist convective
instability into radiative--convective models. These models will require additional computational
resources beyond the already substantial speed and memory requirements of fully non--gray
radiative--convective codes. Perhaps most challenging from a mathematical point of view is to
characterize and incorporate the size--frequency spectrum of broken clouds. The problem is
somewhat akin to the classical one of computing Rosseland mean absorption coefficients in gray
atmosphere models, where the greatest flux contribution comes from the frequency windows of
smallest opacity. Quantifying a broken cloud atmosphere through a single number characterizing
percentage of area covered by clouds is likely to be highly inaccurate, because the amount of
thermal radiation escaping will be highly sensitive to the gaps between the clouds. In the 
near--term, the most promising observations for constraining such a complex situation will be high
spectral resolution data obtaining from large telescopes with sensitive electronic detectors.
However, the theoretical challenge of accurately modeling broken clouds with existing
computational resources remains unsolved.

\mainsection{{V}.~~MODELS OF THE BROWN DWARF GLIESE 229 B}
\backup

The wide gap between stars and brown dwarfs near the edge of the main sequence on the one
hand, and Jupiter on the other hand, is fortuitously occupied by the cool brown dwarf,
Gl 229 B.  It is sufficiently bright to allow high resolution and high signal-to-noise spectroscopy.
This fascinating object displays phenomena common to giant planets and to very-low-mass stars and
represents a benchmark for modeling atmospheres of EGPs.
Model spectra for Gl229 B (Marley \etal 1996; Allard \etal 1996; Tsuji \etal 1996)
reproduce the overall energy distribution fairly well and all agree  that
1) T$_{\rm eff} \sim 950\,$K, 2) the silicate opacity is small compared to
the gaseous molecular opacity and can be ignored in a first approximation,
and 3) the gravity of Gl229 B is poorly constrained at present.
The models, however, fail to reproduce the visible flux, the observed
depth of the strongest molecular absorption bands, as well as the detailed
structure of the observed spectrum.  The calculation of better models for
Gl229 B is currently limited by the inadequate knowledge of CH$_4$ opacities
and a very limited understanding of
the role of grain opacity in such a cool object.

Aside from possible grain absorption and scattering, the spectrum of Gl229 B is shaped
entirely by absorption by H$_2$, H$_2$O, CH$_4$ and, in the mid-infrared,
NH$_3$.  While the opacities of H$_2$ and H$_2$O are now well understood over the entire temperature
range of interest for Gl229 B ($500 \wig< T \wig< 2000\,$K), the current databases for CH$_4$
and NH$_3$  are limited to $T \le 300\,$K (HITRAN database, Rothman \etal 1992) and have an 
incomplete wavelength coverage.
In the case of CH$_4$, this incompleteness is partly remedied by complementing the line list with
frequency-averaged opacity (Karkoschka 1994; Strong \etal 1993).  Because CH$_4$ plays an
important role in shaping the 1 to 4$\,\mu$m spectrum of Gl229 B, the near-infrared fluxes
of the  present models are of limited accuracy, precisely where Gl229 B is brighter and most
easily observable ($Z$, $J$, $H$, and $K$ bands).

The models of Allard \etal (1996) and Tsuji \etal (1996) include TiO and VO opacities (as well
as metal hydrides) which have very strong bands in the red part of the spectrum. As a consequence, their
synthetic spectra reproduce the rapid decrease of the flux shortward of 1$\,\mu$m,  as evidenced by
$R$ and $I$ photometry (Matthews \etal 1996), quite nicely.  However, the visible spectra of Schultz 
\etal (1998) and
Oppenheimer \etal (1998) show a strong H$_2$O band but {\it not} the bands of oxides and hydrides
predicted by those models.  As anticipated by Marley \etal (1996), refractory
elements condense in the atmosphere of Gl229 B and are removed from the gas phase  
(Burrows \& Sharp 1998) and a significant fraction of the condensed particles
may settle to unobservable depths in the atmosphere.
Fluxes from models which do not include grain opacity in the visible predict visible
fluxes which are grossly in error.  

The lack of any other important molecular spectral features in the
visible (Griffith \etal 1998; Saumon \etal 1998) suggests that a source of continuum
opacity is required.  An opacity source that fits these requirements is sub-micron grains.
Mie scattering theory predicts that 0.1 micron radius grains can provide substantial opacity
below about 1 micron yet still be transparent at longer wavelengths (Fig. 2).  
%
%
In fact, such behavior is commonly observed in the solar system and has been the solution for
such diverse puzzles as the radar scattering behavior of Saturn's rings particles and the
spectrum of Venus' atmosphere.

Figure 2 demonstrates the dependence of the opacity of small
particles upon their size.  Submicron particles interact
strongly with optical radiation, while affecting only slightly
longer wavelengths.  Griffith et al. (1998) investigated a range of
possible particle optical properties and sizes and found that
very red, abosorbing, submicron particles can indeed lower the optical flux
of Gliese 229B, while only slightly affecting the near-infrared spectrum.
Griffith \etal speculate that since Gl229 B
receives twice the UV flux
of Titan, an object whose atmosphere is dominated by a photochemical smog
of condensates, the grains in the brown dwarf's atmosphere could
also consist of photochemically--derived non--equilibrium species.
Such carbonaceous material is seen throughout the solar system
and is generally recognizeable by its very red color.

\subsection{A.~~Gravity}

It is highly desirable to constrain the value of the surface gravity of Gl229 B to an astrophysically
useful range.  As reported by Allard \etal (1996), the spectral energy distribution of Gl229 B
models is fairly sensitive to the gravity.  
The most gravity-sensitive colors are $H-K$ and $J-H$ with $\Delta(H-K) / \Delta (\log g)
= -0.39$ and $\Delta(J-K) / \Delta (\log g) = -0.26$, respectively. 
However, the uncertainties in the photometry of Matthews \etal (1996) in these color indices
are comparable to this gravity sensitivity.  The present uncertainties in CH$_4$
opacities and in the role of grains also limit the ability of the models to predict reliable
near-infrared colors.

Nevertheless, it is possible to constrain the gravity by analyzing the spectrum in the 1.9 -- 2.1$\,\mu$m
region (Saumon \etal 1998). In this unique part of the spectrum of Gl229 B, the absorption is completely
dominated by
two well-understood opacity sources:  H$_2$O and the collision-induced absorption by H$_2$.  Methane
absorption is more than two orders of magnitude weaker and unlikely to become significant even when
high-temperature CH$_4$ opacities become available.  The importance of H$_2$O opacity in this region
is confirmed by the remarkable correspondance of the features of the observed spectrum and of
the opacity
of water (Geballe \etal 1996).  The detailed features of the synthetic spectrum are therefore
far more reliable in the 1.9 -- 2.1$\,\mu$m region than in any other part of the spectrum.
%
%
Figure 3 shows how the gravity affects the modeled features of water in this narrow spectral range.
The models shown all have the bolometric luminosity of Gl229 B given by Matthews \etal 1996.
The top curve shows the $\Teff=870$$\,$K, $\log g=4.5$ model.  To emphasize the effects of changing
the gravity, the flux {\it ratio} between this model
and the $\Teff=940$$\,$K, $\log g=5$ model (middle curve) and the $\Teff=1030$$\,$K, $\log g=5.5)$ model 
(lower curve) are shown.  If there were
no gravity sensitivity, the two lower curves would be flat.  This region contains thousands of
H$_2$O features and their number and depth increases with the spectral resolution.  Correspondingly,
the gravity sensitivity washes out at low resolution.  High signal-to-noise spectroscopy at a 
resolution of $R \wig> 1500$ should show this effect very well and
has the potential of reducing the uncertainty in the surface gravity of Gl229 B to $\pm$0.25 dex.

\subsection{B.~~Metallicity}

The metallicity of late M dwarfs is notoriously difficult to determine and modern efforts
are still limited to a small number of stars (see, for example, Leggett \etal 1996, Schweitzer \etal 1996, 
Viti \etal 1997).  
The metallicity of the primary star, Gl229 A, appears to be approximately solar, but is rather uncertain. 
Published values give $[\rm{Fe/H}]=-0.2$ (Schiavon, Barbuy, \& Singh 1997) and [M/H]=0.2 (Mould 1978).
The case of Gl229 B is even more problematic. 
The relative success of the current models indicate that Gl229 B is also approximately solar 
in composition.  However, if the brown dwarf formed from a dissipative accretion disk, it may be analogous 
to the giant planets of the solar system and be enriched in heavy elements.  On the other hand,
phase separations in the interior may deplete the atmosphere of its heaviest elements.

Grain formation complicates the definition of the metallicity of an atmosphere.  Elements
are selectively removed from the gas phase and introduced in condensed phases whose
composition is far more difficult to establish by spectroscopy than that of the gas.
Since the gas-phase abundance of refractory elements is quite sensitive to the physics of
condensation, it is best to focus on the abundant metals C, N, and O which are not
significantly depleted by condensed phases in the atmosphere of Gl229 B.

The limitations of the CH$_4$ opacity database currently prevents a reliable
determination of [C/H] in Gl229 B. While the opacity of H$_2$O, the main oxygen-bearing
molecule, is now well understood,
the metallicity dependence of the synthetic spectra is muddled by the
presence of dust affecting the near infrared.  The need to untangle the
veiling due to dust and the effects of metallicity on the H$_2$O absorption bands make it
very difficult to determine the [O/H] ratio in Gl229 B, although
Griffith \etal (1998) find that a best fit to the optical spectrum
of Gl229 B is achieved with a subsolar oxygen abundance ($[\rm{O/H}]=-0.2$).  

Our model spectra predict a strong feature of NH$_3$ near 10.5$\,\mu$m which has not yet been
detected by observers (Marley \etal 1996, Saumon \etal 1998).  The identification of NH$_3$ 
in the spectrum of Gl229 B would represent
the first detection of this molecule in a compact object outside of the solar system. It also
offers a good possibility of measuring the [N/H] ratio, since dust opacity is
negligible in Gl229 B at this long wavelength.

The model atmospheres of Burrows \etal (1997) all assumed a solar abundance 
of the elements.  In fact, the metallicity of the sun is somewhat higher than that of the 
average star and it is appropriate to consider a greater variety of atmospheric 
metallicities.  We have begun this process by computing atmosphere models for brown 
dwarfs in the mass and temperature range of Gliese 229 B.  These exploratory models were 
constructed by varying the mixing ratio of all molecules uniformly away from that predicted 
for a solar mixture of the elements in thermochemical equilibrium.  In actuality, the relative 
mixing ratios of the molecules will not change uniformly as the overall metallicity changes. 
However, such departures will be slight compared to the large range in overall metallicity 
that we have considered.

Low--resolution spectra for three models, each with T$_{\rm eff}=1000\,\rm K$ and 
$g=1000\,\rm m\,sec^{-2}$, but with varying metallicities, are presented in Fig. 4.  
%
%
The models were computed following the procedures in Burrows \etal (1997), although the 
treatment of Rayleigh scattering has been improved.   In the figure, the metallicities are 
varied over the exceptionally large range of [Fe/H] = $-2$ to 1 to demonstrate the overall 
trends.  Generally, as the metallicity decreases, the temperature profile adjusts. More flux 
emerges in the 3 to $5\,\rm \mu m$ band as the continuum molecular opacity falls.  Surprisingly, 
this redistribution of flux results in a decrease in the flux emerging from the depths of the 
near--infrared water bands.  Instead, the lower molecular opacity in the window regions 
allows more flux to emerge from deeper, hotter layers of the atmosphere resulting in a 
larger flux in the windows, a cooler upper atmosphere, and less flux in the depths of the 
near-infrared water and methane bands.  
Beyond about $2\,\rm\mu m$, the overall flux rises as the metallicity increases.  The larger 
metallicity closes off the near-infrared windows, raising the upper atmosphere temperature, 
and increasing the continuum flux.   

The changes in metallicity considered above produce relatively few changes in the color 
differences of Gl229 B-like models, particularly for solar and subsolar abundances.  
Significant changes are found only for increases in metallicity above solar.  $J-H$
and $J-K$ 
are most sensitive to such metallicity variations, as both colors become redder 
as the metallicity increases above solar.  $H-K$ and $K-L^{\prime}$ are relatively insensitive.

\subsection{C.~~Convection}

The molecules found in the atmosphere of a brown dwarf or EGP constrain 
atmospheric structure, dynamics, and chemistry.  By identifying 
the atmospheric composition, spectroscopy provides information on the physical processes which govern 
the atmosphere.

Departures of atmospheric composition from equilibrium are especially interesting.  CO, 
$\rm PH_3$, $\rm GeH_4$, and $\rm AsH_3$ have all been detected in Jupiter's 
atmosphere at abundances many 
orders of magnitude higher than expected from equilibrium chemistry (see review by Fegley \& Lodders 1994). 
The presence of these non-equilibrium molecules is taken to be evidence of convection.  
Since convective timescales are shorter than chemical equilibrium timescales, these 
molecules can be dredged up from deeper in Jupiter's interior and transported to the visible 
atmosphere.   

In Gl229 B, the detection of CO in abundances in excess of that predicted for chemical 
equilibrium (Noll \etal 1997) implies that the visible atmosphere (near 800 to 1400 
K) must also be convective.  Yet many atmosphere models find that the radiative-convective 
boundary lies far deeper, below 1700 K.  However, the models of Marley \etal (1996) and 
Burrows \etal (1997) predict an additional, detached, upper convection zone.  Such 
a zone would transport CO to the visible atmosphere from depths where it is more abundant.  
Other molecules may also trace convection, including $\rm PH_3$.
The chemical equilibrium profile of Cs is very similar to that of CO (Burrows \& Sharp 1998). 
Thus, the same convection that dredges CO must also be dredging Cs.  However, the lack of 
TiO and other refractory diatomics in the spectrum suggest that the atmosphere is not 
fully convective to the depth (below 2000 K) where these molecules condense.  Taken together, 
these results may support the presence of a detached convection zone.  Thus, CO and Cs may be 
tracing the vertical convective structure of the brown dwarf.  A similar radiative zone, lying 
below Jupiter's visible turbulent atmosphere, has been predicted by Guillot \etal (1994).  A 
confirmation of such a zone at Gl229 B would strengthen the  argument for such a zone in 
Jupiter.  $\rm PH_3$ is also potentially detectable in Gl229 B by space--based platforms  
and will also act as a tracer of convection (Noll \etal 1997).  Measurements of the abundances of 
this suite of molecules in a variety of objects may map out the atmospheric dynamics of 
substellar objects.         

Photochemistry driven by incident radiation can also produce important non--equilibrium 
species.  Thus, many hydrocarbons are found in the atmospheres of 
the solar Jovian planets, including $\rm C_2H_2$ and $\rm C_2H_6$ that would not otherwise be expected.  
A rich variety of photochemical products will 
likely be found in the atmospheres of the extrasolar planets, particularly those with warm 
atmospheres and large incident fluxes.  Hazes produced by the condensation of 
some species can produce signatures in the spectra of these objects far in excess of what 
might be expected given their small mixing ratios.

\mainsection{{V}{I}.~~ALBEDOS AND THE REFLECTIVITY OF EGPS}
\backup

Both scattered  light  from the primary star and thermal emission by a planet contribute to a 
planet's  spectrum.  These two components can be crudely modeled as the sum of the 
reflection of a high temperature Planck function characterizing light from the primary plus a 
second, lower temperature, Planck function representing the thermal emission from the 
planet  itself (Saumon \etal 1996). For the planets of our solar system, the 
two Planck functions are well separated in wavelength.  This can be seen from the Wien 
displacement law, $\lambda T = 0.29\,\rm cm \, deg$.  For a 6000 K primary and a planet 
radiating at 200 K, the Planck functions of the primary and planet peak at 0.48 $\rm \mu m$ and 
$14.5\,\rm \mu m$ respectively.   This separation in the bulk of the radiation field from the 
planets and the Sun has led to a specialized nomenclature in which ``solar" and ``planetary" 
radiation are often treated separately.

However, for an arbitrary planet orbiting at an arbitrary distance from its primary, there can 
be substantial overlap of the two Planck functions.  A general theory of extrasolar 
atmospheres must consistently compute the absorbed and scattered incident light. 
Marley (1998) has generated exploratory EGP atmosphere models
that include deposition of incident radiation.  He finds
that in typical EGP atmospheres absorption in the strong
near--IR water and methane bands produces temperature
inversions above the tropopause, similar to Jupiter's stratosphere.
Generation of a comprehensive suite of model reflected and emitted
spectra will require that a large range of primary stellar types, orbital
distances, and planetary masses and ages be investigated.   

Planets are brightest near the peak in the solar Planck function and the 
reflected flux falls off at shorter and longer wavelengths.  To remove the effects of the solar 
spectrum and more clearly understand the processes acting in the planet's atmosphere, the 
reflected spectra of planets are commonly presented as geometric albedo spectra.  The 
geometric albedo is essentially the planetary spectrum divided by the solar spectrum.  
Formally, it is the ratio of the flux received at Earth at opposition to the flux that would be 
received by a Lambert disk of the same size as the planet at its distance from the sun.  

Reflected planetary spectra can also be approximated
by computing wavelength--dependent geometric albedos of EGP
atmospheres.  The resultant planetary spectrum is then
the sum of the emitted flux plus the product of the
incident radiation times the geometric albedo with a phase
correction.  
%
%
Figure 5 compares the geometric albedo spectra of Jupiter and Uranus.  A purely 
Rayleigh scattering planet would have a geometric albedo of 0.75 at all wavelengths.

As Fig. 5 demonstrates, planets are not gray reflectors.  They reflect best near 
$0.5\,\rm\mu m$ where Rayleigh scattering dominates the reflected flux.  At shorter 
wavelengths, Raman scattering, which shifts some UV photons to longer wavelengths, and 
absorption by the ubiquitous high altitude haze found throughout the outer solar system, 
lower the geometric albedo.  Longward of $0.6\,\rm\mu m$ the strength of molecular 
rotational-vibrational bands increases and molecular absorption, rather than scattering,  
begins to dominate the spectrum.  Methane and the pressure-induced bands of hydrogen are 
the most important absorbers.  In between the molecular bands, solar photons reach bright 
cloud decks and are scattered.  So the planets remain bright in some band passes.  

Marley \etal (1998) computed geometric albedo spectra for a large variety of EGPs, 
ranging from planets of less than 1 Jupiter mass to the most massive brown 
dwarfs.  They considered objects with effective temperatures between 100 and 1200 K  
and found that the UV and optical spectra of extrasolar giant 
planets are generally similar to those of the solar giant 
planets.  At longer wavelengths, however, the reflected flux depends 
critically on the presence or absence of atmospheric condensates. 
When condensates are present, photons have the opportunity to  scatter before 
they are absorbed.   

The most important condensate in EGP atmospheres is water.
Water clouds appear as EGPs cool through effective temperatures
of about 400 K.  The sudden appearance of water clouds
brightens the planets in reflected red and IR light,
as shown in Figure 6.
%
%
This figure presents a computed spectrum for an extrasolar 
planet orbiting at 1 A.U. from a G2V primary.  Also shown is the emitted flux for an object 
with T$_{\rm eff} = 400\,\rm K$.  It is apparent from this plot that the thermal emission 
dominates in the infrared for cloud--free objects.  However, once clouds form they will 
both attenuate the emitted flux and reflect a far larger proportion of the incident radiation.    
Of course, there will be a continuum such that for objects with the same effective 
temperature, but which are closer to their primary, the reflected flux will surpass the emitted flux. 
The emitted flux will continue to be important  
at lower temperatures for objects further from their primaries.  Again, individual models are 
required for each specific case.

The ratio of the total reflected light to the total 
light incident upon a planetary atmosphere (Bond albedo)
depends sensitively on the spectral distribution of the
incident radiation.
For example, most of the flux of 
an A star emerges in the UV and blue.  When incident upon a giant planet, most of this 
light may be Rayleigh scattered, resulting in a large Bond albedo.  In contrast, the 
predominantly red and infrared photons from an M star are far more likely to be absorbed.  
Thus, the Bond albedo of the same planet, when illuminated by two different stars, could  
vary by up to an order of magnitude, as the
stellar type of the primary varies (Marley \etal 1998).
Note that nonequilibrium photochemical hazes can darken the
planets in the UV, further complicating the reflected
spectra and energy budgets.

\mainsection{{V}{I}{I}. THE RELATIONSHIP BETWEEN GIANT PLANETS OF THE SOLAR SYSTEM AND EGPS}
\backup

The advent of the science of EGPs and brown dwarfs, with its many new
discoveries outside the solar system, should not cause us to lose sight
of the central role of our local giant planets as exemplars that can be studied in
unrivaled detail.  There are still many questions that remain concerning Jupiter's 
and Saturn's formation and cooling histories that we must answer before
we can confidently tackle their analogs beyond the solar system.
One of the issues that still surrounds the study of our gas giants is the
character, radial distribution, and origin of their compositions.
Do they have a common formation mechanism with the newly discovered substellar objects,
or is there a distinct mass--related boundary between the formation of such giant planets
($\sim 0.001 M_{\odot}$) and brown dwarfs ($\sim 0.01 M_{\odot}$) (see M. 
Mayor et al., this volume) that is reflected in their compositions?
Jupiter and Saturn provide unique laboratories by which we can
address these and related questions of fundamental interest to planetary
science and astronomy.

\subsection{A.~~Non--Solar Metallicities and Implications for Formation}

The solar system exhibits a progressive change in the bulk composition of
its four giant planets as a function of heliocentric distance.  According to
interior models constrained by improvements in the hydrogen--helium
equation of state and by the Galileo entry probe results for the
abundance of helium in the Jovian atmosphere, Jupiter is close to bulk
solar composition, with a probable enhancement of C, N, and O by about
a factor of $\sim 5$, distributed uniformly in its envelope, and with limits
on a dense central core amounting to $\ge 0.02$ by mass
fraction (or about 6 earth masses maximum) (Guillot, Gautier, and Hubbard,
1997).  Models of Saturn, on the other hand,
are distinctly nonsolar, with envelope enhancements of C, N, and O by about
twice the Jovian factor, and a similar dense core
of $\sim 6$ to 8 earth masses (Guillot et al. 1994).
Uranus and Neptune models seem to resemble a larger version of
the Saturn or Jupiter core
with a thin ($\sim 0.1$ by mass) envelope of hydrogen and
helium (Hubbard, Podolak, and Stevenson 1995).

The traditional interpretation of this sequence, which is now
open to revision in the light of recent detections of extrasolar
giant planets and recent downward revisions of the Jupiter and
Saturn core masses, is that the giant planets Jupiter and Saturn formed by
the capture of nebular hydrogen and helium onto dense Uranus/Neptune--like
cores (Mizuno, Nakazawa, and Hayashi 1978), with subsequent significant
accretion of icy cometesimals being
responsible for enrichment of C, N, and O--bearing molecules in their
envelopes.  Presumably, the capture of nebular hydrogen onto proto--Uranus
and proto--Neptune was less efficient owing to lower nebular density at their
orbital radii and slower accretion of their dense icy cores.

The detection of EGPs at very small orbital radii has strongly
suggested the possibility of significant radial migration of giant
planets during the nebular phase (Trilling et al. 1998).  Recent
observations (Terebey et al. 1998) may support the possibility of direct
formation of giant planets without the necessity of an initial
dense core to trigger hydrodynamic accretion of hydrogen--helium gas.

\subsection{B.~~Constraints on Cooling Mechanisms}

The luminosity of a giant planet is
determined by the heat radiated into space by means
of depletion of its interior entropy as determined by
the atmospheric boundary condition.  The latter is strongly affected
by metallicity.  Moreover, the heat evolved can be strongly increased
by a redistribution of hydrogen (with high specific
entropy) toward shallower, lower--temperature regions of the planet,
while denser components (with lower specific entropy) sink to deeper,
hotter regions.  Thus, atmospheric depletion of metals and helium may
be accompanied by higher luminosity.  It is this effect which is
believed to be responsible for the anomalously high luminosity of
Saturn (Stevenson \& Salpeter 1977).

In principle, the detection of EGPs, whose values of luminosity, age, mass,
and atmospheric composition can be determined, will provide a test of
this concept along with constraints on the hydrogen phase diagram.
The latter predicts that helium separation may be important for
objects with $M < 0.001 M_{\odot}$, while possible segregation of other species
via a first--order phase transition in hydrogen may occur in objects
with masses $M < 0.01 M_{\odot}$.

\mainsection{{} ACKNOWLEDGMENTS}
\backup

We thank David Sudarsky, Christopher Sharp, 
Richard Freedman, Shri Kulkarni, Jim Liebert,
Davy Kirkpatrick, France Allard, Gilles Chabrier, Ben Oppenheimer, 
Chris Gelino, and Tristan Guillot for a variety of useful contributions.
This work was supported under NASA grants NAG5-7499, NAG5-7073, NAG5-4988, and NAG5-2817
and under NSF grant AST93-18970.

\vfill\eject
\null

\vskip .5in
\centerline{\bf REFERENCES}
\vskip .25in

\ref{Allard, F., Hauschildt, P.H., Baraffe, I., \& Chabrier, G. 1996. Synthetic spectra and mass 
determination of the brown dwarf Gl229 B.  {\refit Astrophys.\ J.\ Lett.\/} 465:L123--L127.}

\ref{Allard, F., Hauschildt, P. H., Alexander, D. R. \& Starrfield,
S. 1997. Model atmospheres of very low mass stars and brown dwarfs. 
{\refit Ann.\ Rev.\ Astron.\ Astrophys.\/} 35:137--177.}

\ref{Allard, F. 1998. Model atmospheres: Brown dwarfs from the stellar perspective. In {\refit
Brown dwarfs and extrasolar planets}, ASP conference Series Vol. 134, eds. R. Rebolo, 
E.L. Mart\'in and M.R. Zapatero Osorio (San Francisco: Astronomical Society of the Pacific),
pp.\ 370--382.}

\ref{Baines, K.H., Hammel, H.B., Rages, K.A., Romani, P.N., \& Samuelson, R.E. 1995. Clouds and
hazes in the atmosphere of Neptune. In {\it Neptune and Triton}, (Ed. D.P. Cruikshank). Tucson:
University of Arizona Press, pp. 489--546.}

\ref{Boss, A. 1995.  Proximity of Jupiter-like planets to low-mass stars.
{\refit Science} 267:360--362.}

\ref{Burrows, A., Hubbard, W.B., \& Lunine, J.I. 1989. Theoretical models of vow low mass stars
and brown dwarfs.  {\refit Astrophys.\ J.\/} 345:939--958.}

\ref{Burrows, A., Hubbard, W.B., Saumon, D., \& Lunine, J.I. 1993. An expanded set of 
brown dwarf and very low mass star models.  {\refit Astrophys.\ J.\/} 406:158--171.}

\ref{Burrows, A., Saumon, D., Guillot, T., Hubbard, W.B., \& Lunine, J.I. 1995.
Prospects for detection of extrasolar giant planets by next
generation telescopes. {\refit Nature\/} 375:299--301.}

\ref{Burrows, A., Marley, M.S., Hubbard, W.B., Lunine, J.I., Guillot, T., Saumon, D.,
Freedman, R.S., Sudarsky, D., \& Sharp, C. 1997. A nongray theory of extrasolar giant planets
and brown dwarfs. {\refit Astrophys.\ J.\/} 491:856--875.}

\ref{Burrows, A. \& Sharp, C.M. 1998. Chemical equilibrium abundances in brown dwarf and extrasolar giant
planet atmospheres. submitted to {\refit Astrophys.\ J.\/}.}

\ref{Butler, R. P. \& Marcy, G. W. 1996. A planet orbiting 47 Ursae Majoris. 
{\refit Astrophys.\ J.\/} 464:L153--L156.}

\ref{Butler, R. P., Marcy, G. W., Williams, E., Hauser, H., \& Shirts, P. 1997. Three 
new ``51 pegasi-type'' planets. {\refit Astrophys.\ J.\/} 474:L115--L118.}

\ref{Carlson, R.W., Baines, K.H., Orton, G.S., Encrenaz, Th., Drossart, P., Roos-Serote, M., Taylor,
F.W., Irwin, P., Wier, A., Smith, S., Calcutt, S. 1997. Near-infrared spectroscopy of the
atmosphere of Jupiter. {\refit EOS} 78 supplement:F413.}

\ref{Chabrier, G. \etal 1998. to be published in the proceedings
of the first Euroconference on {\refit Stellar Clusters and Associations},
held in Los Cancajos, La Palma, Spain, May 11--15, eds. R. Rebolo, V. Sanchez--Bejar,
and M.R. Zapatero-Osorio.}

\ref{Cochran, W.D., Hatzes, A.P., Butler, R.P., \& Marcy, G. 1997. The discovery of a 
planetary companion to 16 Cygni B. {\refit Astrophys.\ J.\/} 483:457--463.}

\ref{Emanuel, K.A. 1994. {\refit Atmospheric Convection}. New York: Oxford University Press.}

\ref{Fegley, B. Jr. \& Lodders, K. 1994. Chemical models of the deep
atmospheres of Jupiter and Saturn. {\refit Icarus} 110:117--154.}

\ref{Fegley, B. Jr. \& Lodders, K. 1996. Atmospheric chemistry of the brown dwarf Gliese 229 B: 
Thermochemical equilibrium predictions. {\refit Astrophys.\ J.\ Lett.\/} 472:L37--L39.}

\ref{Geballe, T.R., Kulkarni, S.R., Woodward, C.E., \& Sloan, G.C. 1996. The near-infrared spectrum of
the brown dwarf Gliese 229 B. {\refit Astrophys.\ J.\ Lett.\/} 467:L101--L104.}

\ref{Goody, R.M. \& Yung, Y.L. 1989 {\refit Atmospheric Radiation: Theoretical Basis} New York:
Oxford University Press,  519pp.}

\ref{Griffith, C.A. \etal 1998. in preparation.}

\ref{Guillot, T., Chabrier, G. Morel, P., \& Gautier, D. 1994.
Nonadiabatic models of Jupiter and Saturn.  {\refit Icarus} 112:354--367.}

\ref{Guillot, T. Burrows, A., Hubbard, W.B., Lunine, J.I., \& Saumon, D. 1996. 
Giant planets at small orbital distances.  {\refit Astrophys.\ J.\ Lett.\/}
459:L35--L38.}

\ref{Guillot, T., Gautier, D., \& Hubbard, W.B. 1997.
NOTE: New constraints on the composition of Jupiter from galileo measurements and interior models.
{\refit Icarus} 130:534--539.}

\ref{Hubbard, W.B., Podolak, M., \& Stevenson, D.J. 1995.
The Interior of Neptune.  In {\refit Neptune and Triton}.
ed. D.P. Cruikshank (Tucson: University of Arizona Press),
pp. 109--138.}

\ref{Jones, H.R.A. \& Tsuji, T. 1997. Spectral evidence for dust in late--type M dwarfs.
{\refit Astrophys.\ J.\/} 480:L39--L41.}

\ref{Karkoschka, E. 1994. Spectrophotometry of the Jovian planets and Titan at 300 to 1000 nm
wavelength: The methane spectrum. {\refit Icarus} 111:174--192.}

\ref{Kirkpatrick, J.D., Reid, I.N., Liebert, J., Cutri, R.M., Nelson, B.,
Beichman, C.A., Dahn, C.C., Monet, D.G., Skrutskie, M.F., \& Gizis, J. 1998.
in preparation.}

\ref{Latham, D.~W., Mazeh, T., Stefanik, R.P., Mayor, M., \& Burki, G. 1989.
The unseen companion of HD114762 - A probable brown dwarf. {\refit Nature} 339:38--40.}

\ref{Leggett, S.K., Allard, F., Berriman, G., Dahn, C.C., \& Hauschildt, P.H. 1996.  Infrared spectra
of low-mass stars: Towards a temperature scale for red dwarfs. {\refit Astrophys.\ J.\ Suppl.\/}
104:117--143.}

\ref{Lunine, J.I., Hubbard, W.B., \& Marley, M. 1986. Evolution and infrared spectra of brown
dwarfs. {\refit Astrophys.\ J.\/} 310:238--260.}

\ref{Lunine, J.I., Hubbard, W.B., Burrows, A. S., Wang, Y.-P., \& Garlow, K. 1989.  The effect of
gas and grain opacity on the cooling of brown dwarfs.  {\refit Astrophys.\ J.\/} 338:314-337.}

\ref{Marcy, G. W. \& Butler, R. P. 1996. A planetary companion to 70 Virginis. 
{\refit Astrophys.\ J.\/} 464:L147--L151.}  

\ref{Marley, M.S., Saumon, D., Guillot, T., Freedman, R.S., Hubbard, W.B., Burrows, A.,
\& Lunine, J.I. 1996. Atmospheric, evolutionary, and spectral models of the brown dwarf
Gliese 229 B. {\refit Science} 272:1919--1921.} 

\ref{Marley, M. 1998. Atmosphers of giant planets from Neptune to Gliese 229 B. In {\refit
Brown dwarfs and extrasolar planets}, ASP conference Series Vol. 134, eds. R. Rebolo,
E.L. Mart\'in and M.R. Zapatero Osorio (San Francisco: Astronomical Society of the Pacific),
pp.\ 383--393.}  

\ref{Marley, M.S., Gelino, C., Stephens, D., Lunine, J.I., \& Freedman, R. 1998. 
Reflected spectra and albedos of extrasolar giant planets I: Clear and cloudy atmospheres.
submitted to  {\refit Astrophys.\ J.\/}.}

\ref{Matthews, K., Nakajima, T., Kulkarni, S.R., \& Oppenheimer, B.R. 1996. Spectral energy distribution 
and bolometric luminosity of the cool brown dwarf Gliese 229B. {\refit Astron.\ J.\/} 112:1678--1682.}

\ref{Mayor, M. \& Queloz, D. 1995. A jupiter-mass companion to a solar-type star.  
{\refit Nature} 378:355--357.}

\ref{Mizuno, H., Nakazawa, K, \& Hayashi, C. 1978.
Instability of a gaseous envelope surrounding a planetary core and
formation of giant planets.  {\refit Prog. Theor. Phys} 60:699-710.}

\ref{Mould, J.R. 1978. Infrared spectroscopy of M dwarfs. {\refit Astrophys.\ J.\/} 226:923--930.}

\ref{Nakajima, T., Oppenheimer, B.R., Kulkarni, S.R., Golimowski, D.A., Matthews, K.,
\& Durrance, S.T. 1995. Discovery of a cool brown dwarf. {\refit Nature} 378:463--465.}

\ref{Noll, K.S., Geballe, T.R., \& Marley, M.S. 1997. Detection of abundant carbon monoxide in the
Brown dwarf Gl229 B. {\refit Astrophys.\ J.\ Lett.\/}, 489:L87--L90.}

\ref{Noyes, R.W. \etal\ 1997. A planet orbiting the star Rho Coronae Borealis.
{\refit Astrophys.\ J.\ Lett.\/} 483:L111-L114.}

\ref{Oppenheimer, B.R., Kulkarni, S.R., Matthews, K., \& Nakajima, T. 1995. Infrared spectrum of the cool
brown dwarf Gl229 B. {\refit Science} 270:1478--1479.}

\ref{Oppenheimer, B.R., Kulkarni, S.R., Matthews, K., \& van Kerkwijk, M.H. 1998.
The spectrum of the brown dwarf Gliese 229 B. {\refit Astrophys.\ J.\/} 502:932--943.}

\ref{Orton, G. and 40 others. 1996. Earth-based observations of the Galileo probe entry site. {\refit
Science} 272:839--840.}

\ref{Ragent, B., Colburn, D.S., Avrin, P., Rages, K.A. 1996. Results of the Galileo Probe
nephelometer experiment. {\refit Science} 272:854--856.}

\ref{Rossow, W.B. 1978. Cloud microphysics: Analysis of the clouds of Earth, Venus, Mars and
Jupiter. {\refit Icarus} 36:1--50.}

\ref{Rothman, L.S. \etal 1992. The HITRAN molecular database - Editions of 1991 and 1992.
{\refit J.\ Quant.\ Spectros.\ Rad.\ Transf.\/} 48:469--507.}

\ref{Saumon, D., Hubbard, W.B., Burrows,
A., Guillot, T., Lunine, J.I., \& Chabrier, G. 1996. 
A theory of extrasolar giant planets.  {\refit Astrophys.\ J.\/} 460:993--1018.}

\ref{Saumon, D., Marley, M.S., Guillot, T., \& Freedman, R.S. 1998. Spectral diagnostics for the
brown dwarf Gliese 229 B, submitted to {\refit Astrophys.\ J.\/}.}

\ref{Schweitzer, A., Hauschildt, P.H., Allard, F., \& Basri, G. 1996. Analysis of Keck high resolution
spectra of VB10. {\refit Mon.\ Not.\ Roy.\ Astron.\ Soc.\/} 283:821--829.}

\ref{Schiavon, R.P., Barbuy, B., \& Singh, P.D. 1997. The FeH Wing-Ford band in spectra of M stars. 
{\refit Astrophys.\ J.\/} 484:499--510.}

\ref{Schultz, A.B., \etal 1998. First results from the Space Telescope Imaging Spectrograph: Optical
spectra of Gl229 B. {\refit Astrophys.\ J.\ Lett.\/} 492:L181--L184.}

\ref{Stevenson, D.J. \& Salpeter, E.E. 1977.  The dynamics and helium distribution in hydrogen-helium fluid
planets. {\refit Astrophys.\ J.\ Suppl.\/} 35:239--261.}

\ref{Strong, K., Taylor, F.W., Calcutt, S.B., Remedios, J.J., \& Ballard, J. 1993. Spectral parameters
of self  and hydrogen broadened methane from 2000 to 9500 cm$^{-1}$ for remote sounding of the
atmosphere of Jupiter. {\refit J.\ Quant.\ Spectrosc.\ Radiat.\ Transfer\/} 50:363--429.}

\ref{Terebey, S. et al. 1998. submitted to Nature.}

\ref{Trilling, D.E., Benz, W., Guillot, T., Lunine, J.I.,
Hubbard, W.B., \& Burrows, A. 1998.  Orbital evolution and migration of giant planets: modeling extrasolar
planets. {\refit Astrophys.\ J.\/} 500:428--439.}

\ref{Tsuji, T., Ohnaka, K., Aoki, W. \& Nakajima, T. 1996. Evolution of dusty photospheres through red 
to brown dwarfs: how dust forms in very low mass objects.  {\refit Astron.\ Astrophys.\ Lett.\/} 308:L29--L32.} 

\ref{Viti, S., Jones, H.R.A., Schweitzer, A., Allard. F., Hauschildt, P.H., Tennyson, J., Miller, S.,
\& Longmore, A.J. 1997. The effective temperature and metallicity of CM Draconis. 
{\refit Mon.\ Not.\ Roy.\ Astron.\ Soc.\/} 291:780--796.}

\ref{Yair, Y., Levin, Z., \& Tzivion, S. 1995. Microphysical processes and dynamics of a Jovian
thundercloud. {\refit Icarus} 114:278--299.}


\vfill\eject
\null

\vskip .5in
\centerline{\bf FIGURE CAPTIONS}
\vskip .25in

\caption{
\caption{Figure 1.\capskip Evolution of the luminosity (in L${_\odot}$) of solar--metallicity M dwarfs and substellar objects
versus time (in years) after formation (from Burrows et al. 1997).
The stars, ``brown dwarfs'' and ``planets'' are shown as solid, dashed, and dot--dashed
curves, respectively.
In this figure, we arbitrarily designate as ``brown dwarfs'' those objects that burn deuterium,
while we designate those that do not as ``planets.''
The masses in M${_\odot}$ label most of the curves, with the lowest three
corresponding to the mass of Saturn, half the mass of Jupiter, and the mass of Jupiter.}

\caption{Figure 2.\capskip Approximate Mie scattering efficiency
of various-sized enstatite dust grains.  Submicron
grains can produce large scattering and extinction
opacity at optical wavelengths and yet remain essentially
invisible at longer wavelengths.  However, because they
scatter almost conservatively in the optical, silicates
are poor candidates for the grains responsible for
the lower optical flux of Gl229 B.}

\caption{Figure 3.\capskip Gravity dependence of the H$_2$O features in the 1.9 to 2.1 micron region.
The three synthetic spectra correspond to models which all have the bolometric luminosity of Gl229 B.
The top curve shows the $\Teff=870\,$K, $\log g=4.5$ spectrum, used as a reference spectrum in this
figure.  The other curves are flux {\it ratios}
of the $\Teff=940\,$K, $\log g=5$ spectrum to the reference spectrum, and the ratio of the
$\Teff=1030\,$K, $\log g=5.5$ spectrum to the reference spectrum, shown by the middle and bottom
curves, respectively.  All spectra have been shifted vertically for clarity.
The synthetic spectra are shown at a spectral resolution of 1500.}

\caption{Figure 4.\capskip Model thermal emission spectra
for model atmospheres ($g=1000\,\rm m\,sec^{-2}$,
$T_{\rm eff} = 1000\,\rm K$) of varying metallicity.  These
low-resolution spectra demonstrate that
broad band near--IR flux measurements will
not be able to distinguish differences in
metallicity below about about 1/10 solar.  For
larger metallicities, $J$ band is most sensitive
to increasing metallicity.}

\caption{Figure 5.\capskip Geometric albedo spectra for Uranus, Jupiter,
and a model 7--Jupiter mass planet with $T_{\rm eff}=400\,\rm K$.  The model
includes no clouds, and is thus darker than either planet longward of 
about $0.6\,\rm \mu m$.  Absorption by methane and water removes
incident photons before they can be Rayleigh scattered in the 
model.  In contrast, Jupiter's and Uranus' cloud decks reflect brightly
in between strong methane absorption bands.  Jupiter is much darker
than either the model or Uranus at blue and UV wavelengths.  Dark
photochemical hazes are predominantly responsible for lowering
the reflectivity of the planet in this spectral range.}

\caption{Figure 6.\capskip Model spectra for $12\,\rm M_J$,
$T_{\rm eff} = 300\,\rm K$ planet orbiting at 1 A.U. from a G2V star.
Solid line shows reflected flux received at 10 pc if
planet's atmosphere is cloud-free.  Dotted line
demonstrates enhancement in flux when water clouds
are added to the model.  Both models
are from Marley et al. (1998). Long dashed line gives
thermal emission in the case of no clouds 
(Burrows et al. 1997).  For this
object, reflected near-IR flux begins to dominate thermal
emission when clouds form as the object cools through about
400 K.}

\vfill\eject
\vskip3.5truein
\hbox to\hsize{\hfill\includegraphics{figure1.ps}\kern+0in\hfill}

\vfill\eject
\vskip3.5truein
\hbox to\hsize{\hfill\includegraphics{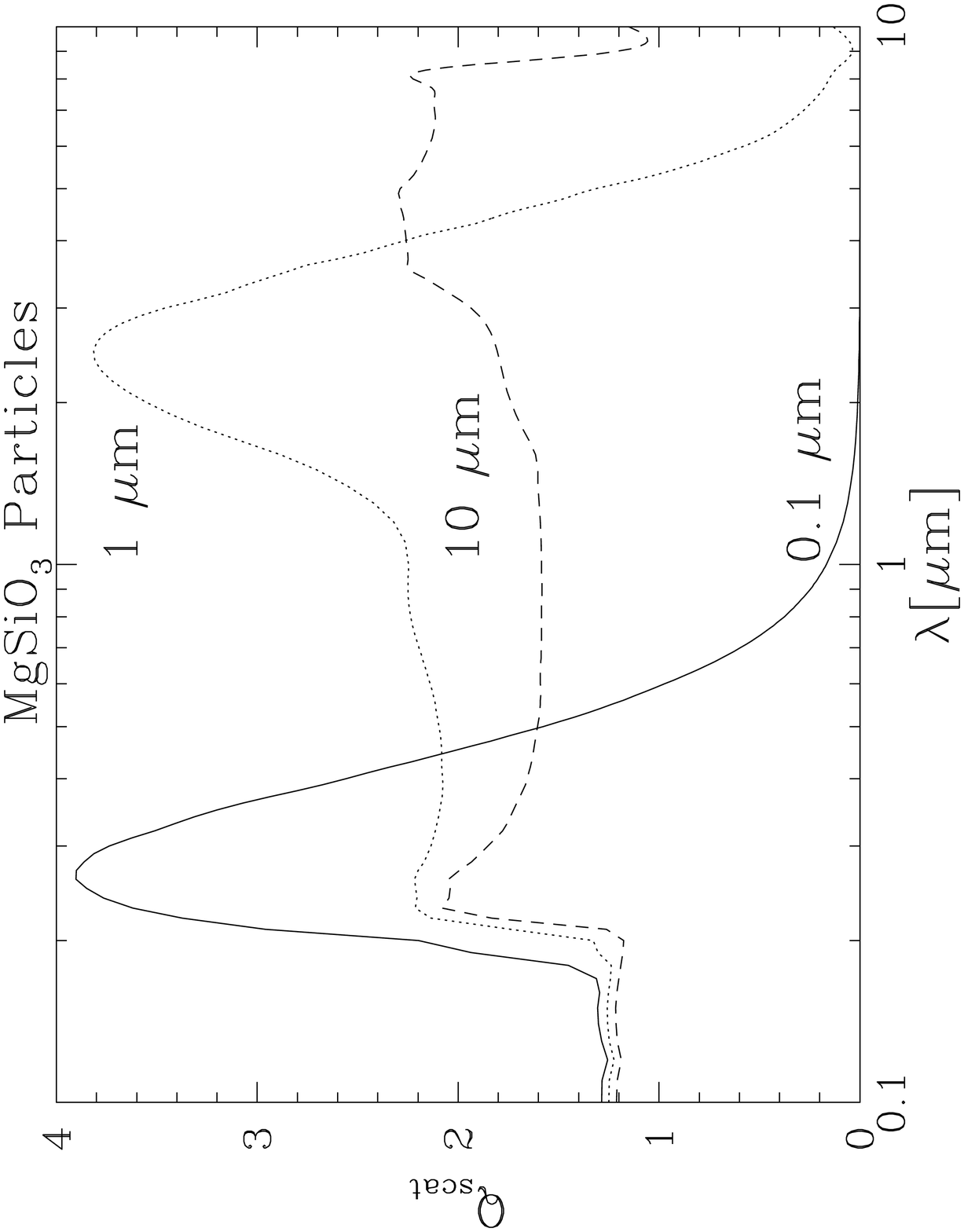}\kern+0in\hfill}

\vfill\eject
\vskip3.5truein
\hbox to\hsize{\hfill\includegraphics{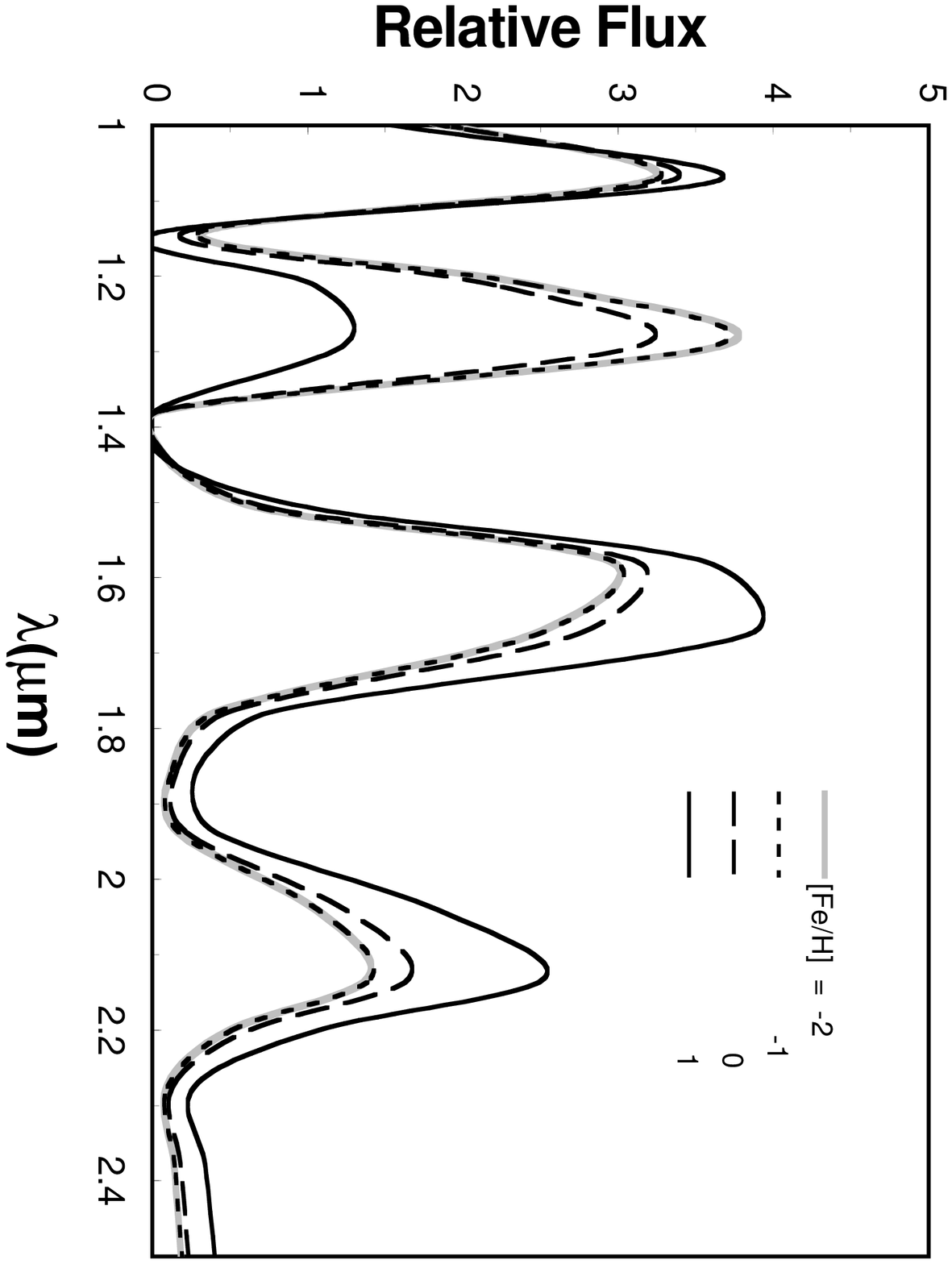}\kern+0in\hfill}

\vfill\eject
\vskip3.5truein
\hbox to\hsize{\hfill\includegraphics{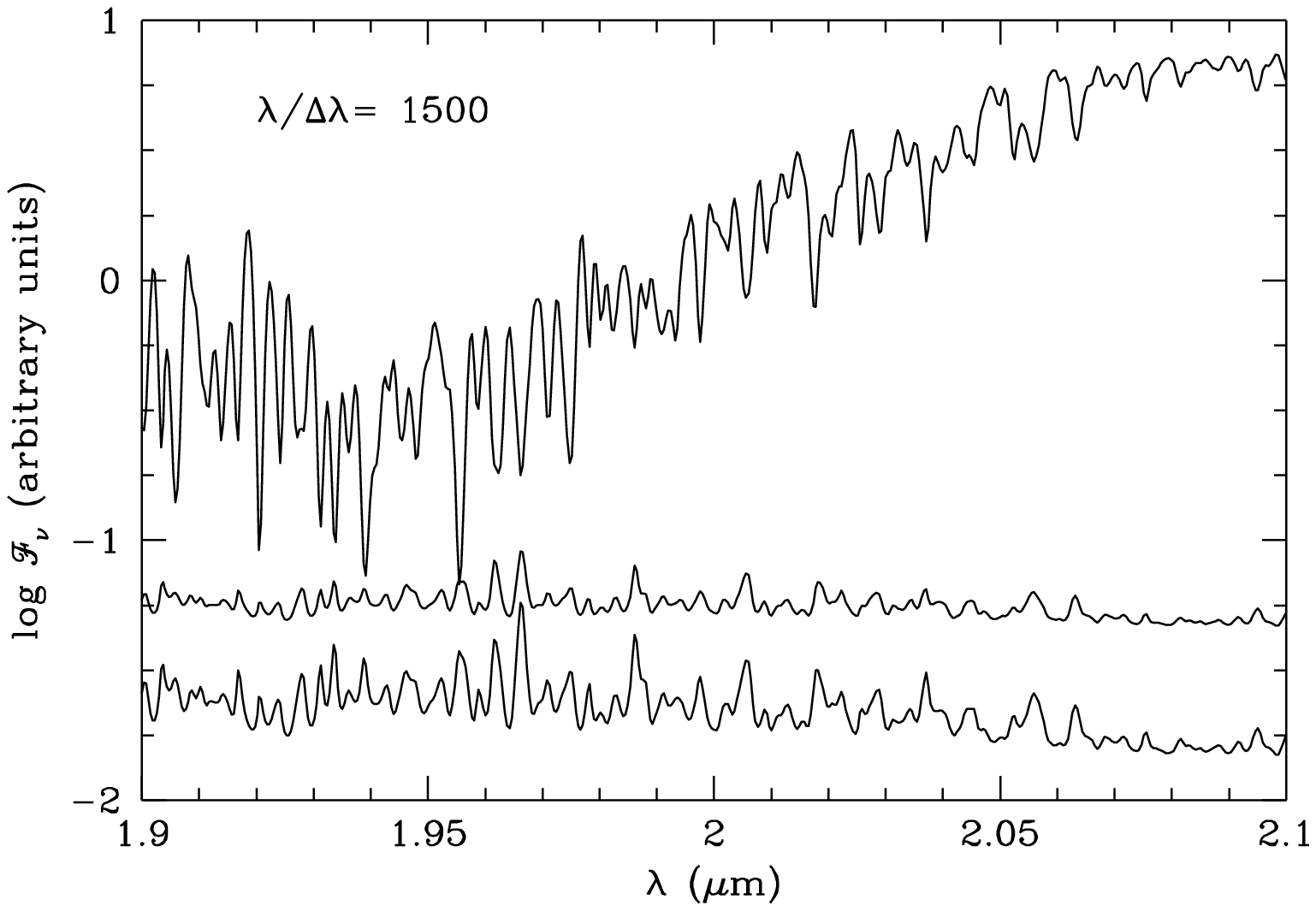}\kern+0in\hfill}

\vfill\eject
\vskip3.5truein
\hbox to\hsize{\hfill\includegraphics{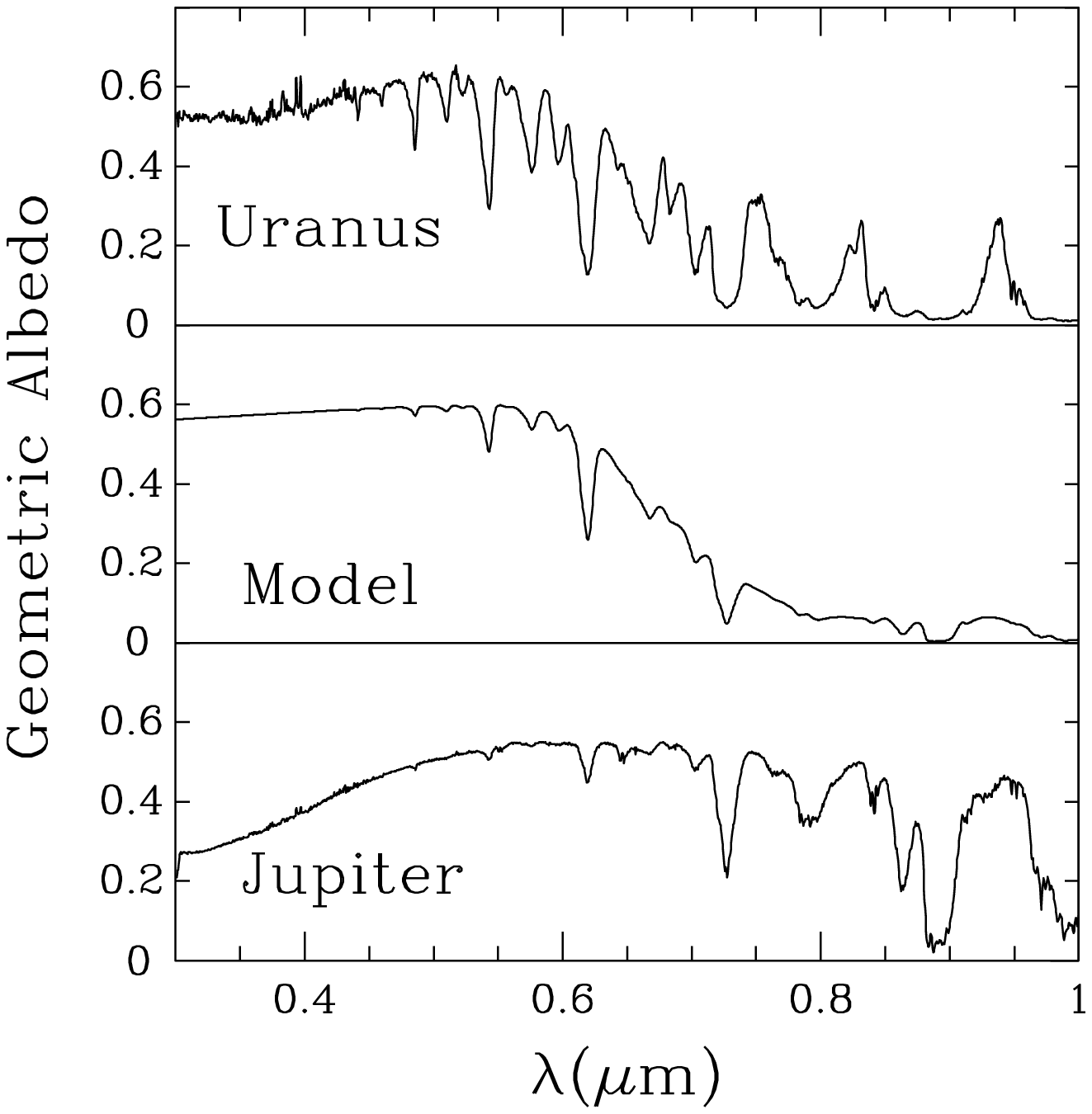}\kern+0in\hfill}

\vfill\eject
\vskip3.5truein
\hbox to\hsize{\hfill\includegraphics{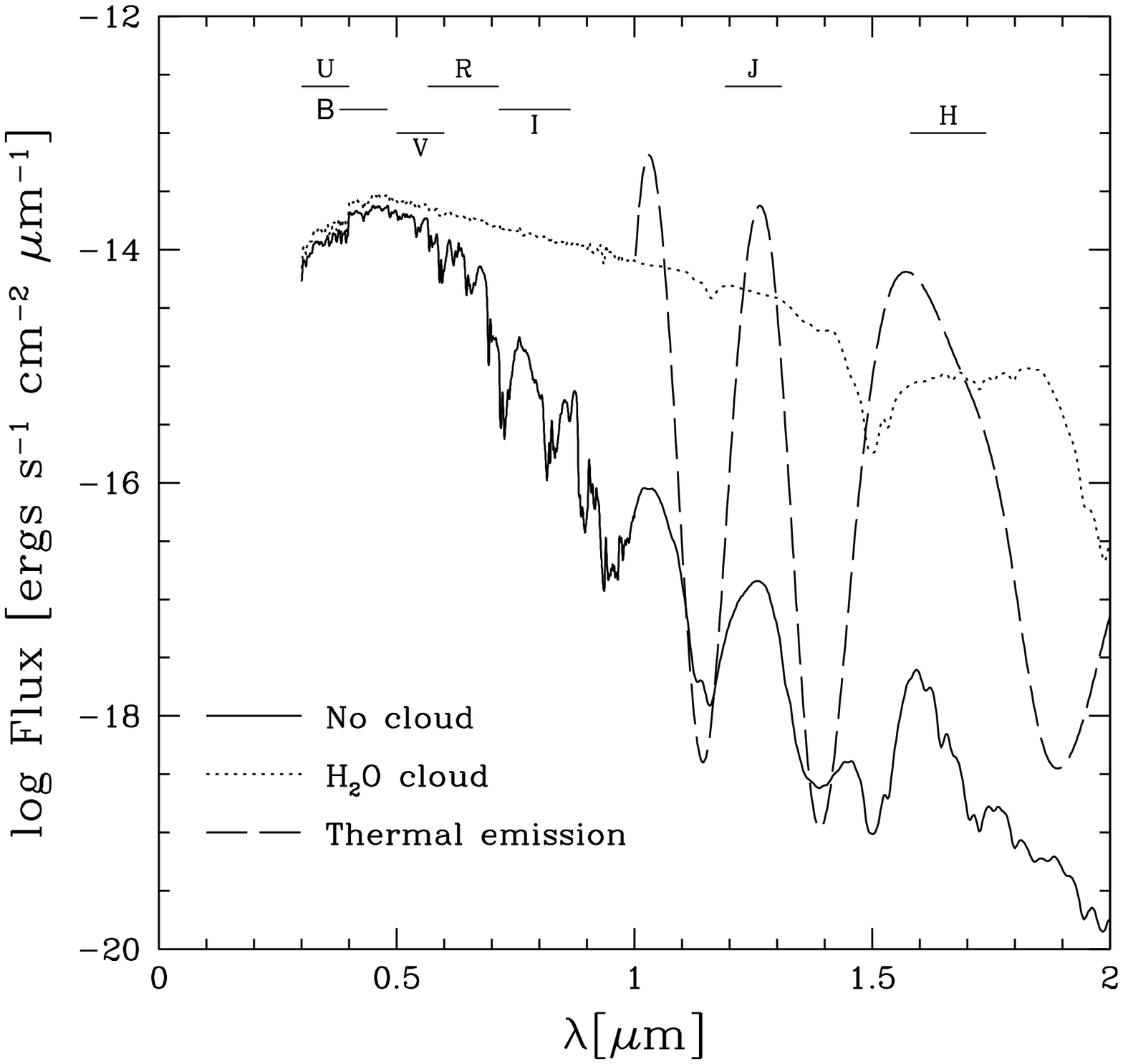}\kern+0in\hfill}

\bye